# Ultrafast electronic read-out of diamond NV centers coupled to graphene

Andreas Brenneis[1,2], Louis Gaudreau[3], Max Seifert[1], Helmut Karl[4], Martin S. Brandt[1], Hans Huebl[2,5], Jose A. Garrido[1,2], Frank H.L. Koppens[3*], Alexander W. Holleitner[1,2*]

[1] Walter Schottky Institut and Physik-Department, Technische Universität München, Am Coulombwall 4a, 85748 Garching, Germany.
[2] Nanosystems Initiative Munich (NIM), Schellingstr. 4, 80799 Munich, Germany.
[3] ICFO - The Institute of Photonic Sciences, Mediterranean Technology Park, Av. Carl Friedrich Gauss 3, 08860 Castelldefels (Barcelona), Spain.
[4] Institute of Physics, University of Augsburg, 86135 Augsburg, Germany
[5] Walther-Meißner-Institut, Bayerische Akademie der Wissenschaften, Garching, Germany

**Nonradiative transfer processes are often regarded as loss channels for an optical emitter[1], since they are inherently difficult to be experimentally accessed. Recently, it has been shown that emitters, such as fluorophores and nitrogen vacancy centers in diamond, can exhibit a strong nonradiative energy transfer to graphene[2–6]. So far, the energy of the transferred electronic excitations has been considered to be lost within the electron bath of the graphene. Here, we demonstrate that the transferred excitations can be read-out by detecting corresponding currents with picosecond time resolution[7,8]. We electrically detect the spin of nitrogen vacancy centers in diamond electronically and control the nonradiative transfer to graphene by electron spin resonance. Our results open the avenue for incorporating nitrogen vacancy centers as spin qubits into ultrafast electronic circuits and for harvesting non-radiative transfer processes electronically.**

With the advancement of nanoscale photonics research, it has become increasingly desirable to combine optical systems with electric circuits to create optoelectronic devices that can be miniaturised and integrated into chips. To this end, we can take advantage of the excellent optical and electronic properties of graphene[9,10], which include good photodetection capabilities[8,11–17], efficient energy absorption[3], and strong light-matter interactions at the nanoscale [18–20]. In particular, it has been recently reported that due to graphene's specific properties the near-field interaction between light emitters and graphene is greatly enhanced as compared to conventional metals[2–6]. This interaction manifests itself e.g. in a hundred-fold enhancement of the excited state decay rate of emitters placed 5 nm away from graphene as compared to the spontaneous emission of the emitter. The physical mechanism behind the interaction is the creation of an electron-hole pair in graphene through nonradiative energy transfer (NRET) from the emitter dipole. The NRET process to graphene has been demonstrated to have an efficiency of nearly 100% when the emitter is less than 10 nm away from the graphene sheet[3], making graphene an ideal material to electrically detect the optical properties of nearby emitters[6]. NRET has been studied extensively for fundamental as well as for biosensing applications. However, fast energy transfer has not yet been observed due to quenching of the optical signal for short graphene-emitter distances. In contrast, an electronic read-out of the NRET enables studies on fast energy processes. Moreover, if the transferred energy can be collected, as we show in this work, new ways for energy harvesting and biosensing can be implemented.

We take advantage of the highly efficient NRET process to electronically read-out, for the first time, the optical excitation of nitrogen vacancy centers (NV centers) in diamond nanocrystals. To this end, we use graphene for extraction of the excited state energy of the NV centers and convert it to a measurable electrical signal. We choose NV centers as optical emitters due to their outstanding characteristics[21] in terms of robustness, stability, spin photon coupling, ease of spin manipulation[22], and spin coherence, which have allowed them to play a fundamental role in new quantum technologies: room temperature quantum registers based on the NV spin[23–25], spin-spin entanglement[26,27], spin-photon entanglement[28], and single photon emitters[29], amongst others. Besides quantum information processing, NV centers have also been used in metrology applications as extremely sensitive nanoscale magnetometers[30] and thermometers[31–33]. These applications require high photon

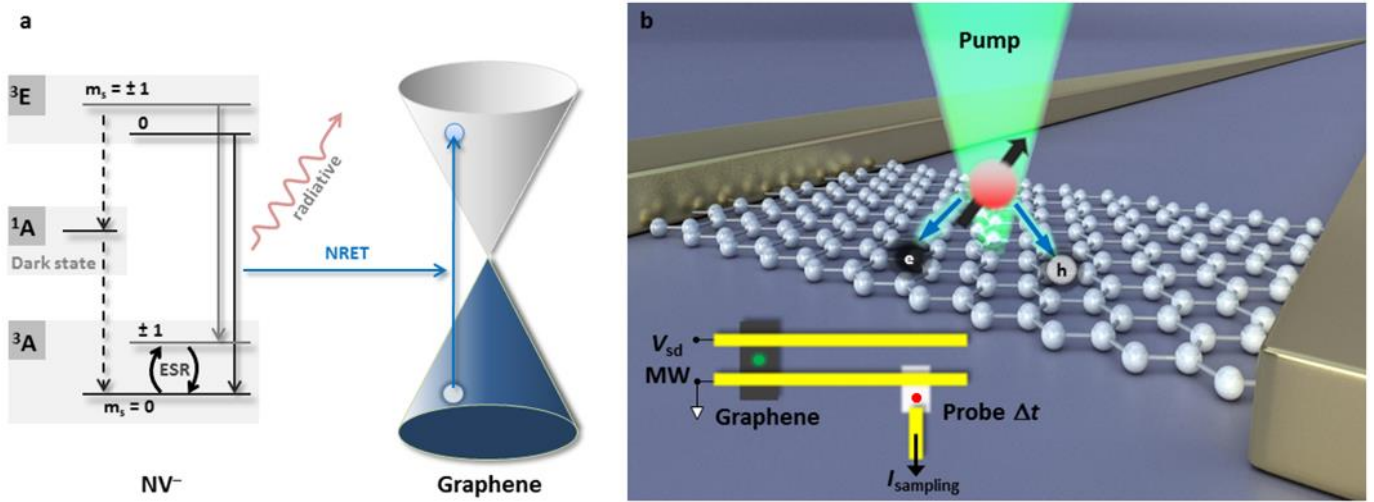

**Figure 1 a,** The optically excited state of the NV center with $m_s = 0, \pm 1$ allows a nonradiative energy transfer (NRET) to the graphene (blue horizontal arrow). The optical excitation of the NV center is spin conserving. However, the NV center comprises a spin selective metastable dark state $^1A$, which offers an additional decay channel for $m_s = \pm 1$ (dashed arrows). The $^1A$ state can thus reduce the NRET rate for the state $m_s = \pm 1$ compared to the state $m_s = 0$. **b,** Excitations caused in the NV center (red sphere with black arrow) can be transferred to the graphene via the NRET (blue arrows) and thus contribute to the current in the graphene. Two striplines act as near-field antenna and propagate all currents as electromagnetic transients (inset). The transients are sampled at an ultrafast Auston switch with a time delayed probe pulse (red circle). Thus, we can read out a dc current $I_{sampling}(\Delta t)$ for a fixed time delay $\Delta t$ between the pump- and the probe-pulse. The microwave (MW) for the electron spin resonance (ESR) is applied to the striplines.

collection efficiencies and therefore bulky collection optics. Hence, an alternative method to extract the spin information from the NV centers is desirable. So far, the spin dynamics of NV centers have only been detected optically via fluorescence measurements. In this letter, we demonstrate ultrafast on-chip electronic detection of NV spins.

Fig. 1a shows the NRET process schematically. Upon optical excitation of the NV center, the relaxation process occurs via NRET into the graphene generating an electron-hole pair. We detect the corresponding electronic excitations originating from the NRET process by utilizing an Auston switch[7,8] that enables an on-chip ultrafast electronic read-out with picosecond resolution (inset Fig. 1b, Methods). The measured current can be understood by considering the extra electron-hole pairs in the graphene, which are generated with a time rate corresponding to the lifetime of the NV centers modified by the NRET. In order to verify that the current observed is due to the nonradiative loss channel of the NV centers, we use the electron spin resonance (ESR) of the NV centers to modify the rate of NRET[34]. This change in rate induces a spin-dependent variation of the current in graphene.

Our device is configured as follows: A graphene sheet is coated with randomly distributed nanodiamonds (Fig. 2a), each containing ≈ 500 NVs (Methods). Two gold striplines serve as electronic contacts to the graphene. We characterize the nonradiative transfer dynamics by measuring the current through the Auston switch while activating it with a laser probe pulse at a time $\Delta t$ after a laser pump pulse in the graphene region (inset of Fig. 1b and Methods)[8]. The circles in Fig. 2a highlight two positions where nanodiamond clusters lie on top of graphene. At such positions, we detect a time-resolved current lasting for 100s of picoseconds (blue shaded part in Fig. 2b and supplementary Figs. S1 and S2). Fig. 2c depicts a time-resolved current trace measured close to one of the metal contacts (triangle in Fig. 2a). At this position, the electric field is dominated by the built-in potential at the graphene-gold interface[8,15,16,35]. The long-lived, transient currents (blue shaded region in Fig. 2b and 2c) are caused by drift and diffusion of electron-hole pairs generated by NRET from the NV centers and by direct optical absorption within the graphene. The excited electron-hole pairs directly generate a photoresponse[11] or decay into hot electrons and generate a photoresponse by the Seebeck effect[36]. The first peak at a time delay of $\Delta t \approx 0$ ps corresponds to an ultrafast displacement current within the graphene; i.e. the screening of the local electric field caused by photogenerated charge carriers[7,8]. For each position, we plot the long-lived, transient currents for increasing time frames (Figs. 2d, 2e, and 2f). We detect such long-lived currents at positions where clusters of nanodiamonds are located (e.g. circles and triangle). The long-lived currents have a timescale between picoseconds up to nanoseconds, which exceeds the direct laser generated carrier dynamics in pristine graphene[8,15,16,35] (supplementary Fig. S3). In addition, the measured decay time is significantly shorter than the natural, characteristic lifetime of NV centers[21]. Therefore, these are signatures that the long-lived currents comprise the nonradiative transfer dynamics.

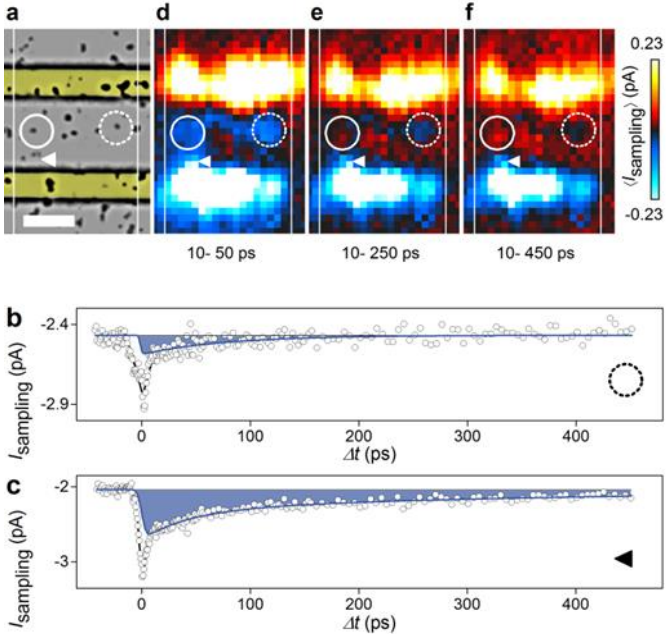

**Figure 2 Ultrafast electronic read-out of NV centers. a,** Graphene (in-between the white vertical lines) with diamond nanocrystals on top. Two metal striplines (yellow) act as source-drain contacts. Scale bar, 10 μm. **b,** and **c,** Time-resolved current $I_{sampling}$ at positions marked by dashed circle and a triangle in Fig. 2a, respectively. **d, e,** and **f,** Color maps of the averaged long-lived currents as highlighted by blue areas in Figs. 2b and 2c for time intervals of $\Delta t$ for $(10-50)$ ps (2d), $(10-250)$ ps (2e), and $(10-450)$ ps (2f). Experimental parameters are 77 K, $P_{laser} = 3.0$ mW, $V_{sd} = 0$ V.

In order to verify that that the long-lived contributions to $I_{sampling}$ stem from charge carriers from the NRET process, instead from direct laser light absorption, ESR provides a selective tool. To this end, we apply a microwave signal to the striplines (inset of Fig. 1b and Methods). The corresponding oscillating magnetic field can be tuned to the spin splitting energy of the NV centers with a resonance at $f_{MW} = 2.875$ GHz (ESR in Fig. 1a)[22]. The energy spectrum of the NV center contains a spin triplet system with a $m_s = 0$ state and $m_s = \pm 1$ degenerated states. Upon laser excitation the spin state of the system is conserved[21]. If the NV center was in the $m_s = 0$ excited state, it decays radiatively with a characteristic lifetime of 10 ns to the $m_s = 0$ ground state. Conversely, if the NV center was in the excited state $m_s = \pm 1$, it decays via a dark state to the ground state $m_s = 0$ (Fig. 1a)[21]. The dark state has a lifetime in the order of hundreds of ns. This difference in decay rates between the $m_s = 0$ and $m_s = \pm 1$ excited states is mapped via NRET onto the graphene by the creation of an electron-hole pair with corresponding probability.

Fig. 3a and 3b show $I_{sampling}$ for non-resonant and resonant ESR conditions. For analysis, we fit the data with three contributions: a baseline $I_{offset}$, a Gaussian immediate response $I_{displacement}$ (red line) and an exponentially decaying function $I_{hot}$ (blue line and supplementary Fig. S2). $I_{hot}$ describes the long-lived currents. We find that both $I_{offset}$ and $I_{hot}$ reduce under the spin resonance conditions, while $I_{displacement}$ stays constant. The latter means that the local electric field at this position within the graphene and the displacement of the field are neither altered by the NRET nor the microwaves. The component $I_{hot}$ comprises transfer processes with a time-scale of several picoseconds, while $I_{offset}$ corresponds to a time-scale of $\approx 13$ ns (the inverse of the repetition frequency of the laser 76 MHz). When this offset component is plotted vs. microwave frequency and aliases of the laser-repetition are subtracted (supplementary information), the resulting $I^*_{offset}$ exhibits a clear dip at the ESR frequency (Fig. 3c and supplementary Fig. S4) which is also observed for the component $I_{hot}$ of the current (Fig. 3d). This is consistent with standard ESR frequency-dependent photoluminescence measurements made on the same sample (Fig. 3e). The correspondence between the ESR dip measured optically and electrically is another clear demonstration of the NRET between the NV center and the graphene. The NRET excites electron-hole pairs in the graphene, and therefore, the current is temporarily increased in graphene. At the spin resonance, the relaxation channel through the dark state reduces the NRET, therefore, $I_{hot}$ is reduced. Additional evidence to prove that the ESR signal originates from the NV centers is given in Fig. 3f, where upon excitation below the optical transition energy of NV centers, no ESR dip is observed.

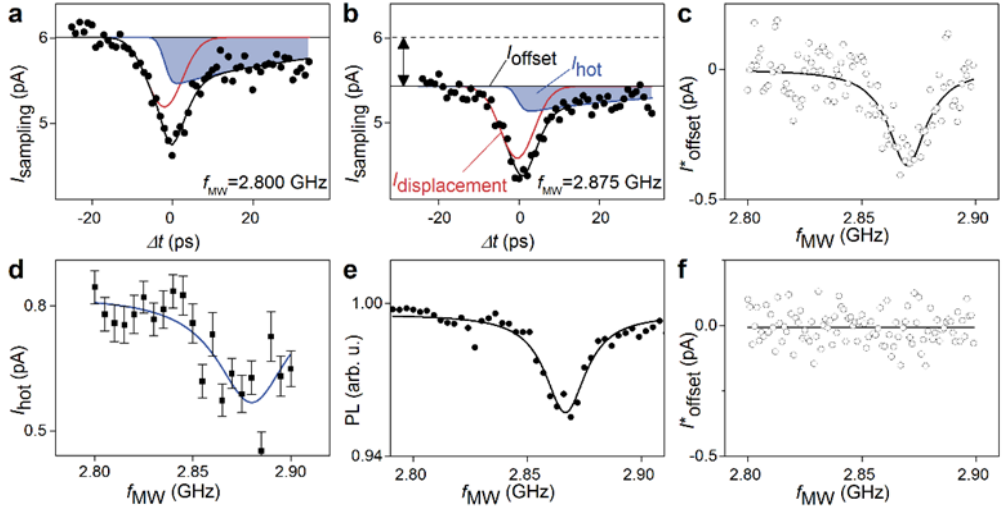

**Figure 3 Electronically and optically detected electron spin resonance. a,** Time-resolved photocurrent under non-resonant microwave irradiation excited at a position of nanodiamond clusters on the graphene. **b,** The signal can be characterized by an offset $I_{offset}$ (horizontal line), a fast peak $I_{displacement}$ (red) and an exponentially decaying function $I_{hot}$ (blue). $I_{offset}$ and $I_{hot}$ reduce under resonant microwave irradiation. The double arrow describes the ESR-reduction of $I_{offset}$ with respect to a non-resonant excitation (dashed line). **c, d,** and **e,** $I^*_{offset}$, $I_{hot}$, and the photoluminescence vs. the microwave frequency $f_{MW}$ with a dip at the spin resonance frequency, respectively. **f,** The ESR signal of the current vanishes for a laser energy (1.53 eV/ 811 nm) below the optical excitation energy of NV center. The experimental parameters are for a and b: $\lambda = 535$ nm, $P_{laser} = 2.0$ mW, $T_{bath} = 77$ K, $V_{sd} = 0.233$ V, $P_{MW} = 1$ W; for c and d: $\lambda = 535$ nm, $P_{laser} = 0.5$ mW, $T_{bath} = 77$ K, $V_{sd} = 0.233$ V, $P_{MW} = 1$ W; for e: $\lambda = 532$ nm, $P_{laser} = 2.0$ µW, $T_{bath} = 300$ K, $P_{MW} = 0.1$ W; for f: $\lambda = 811$ nm, $P_{laser} = 0.7$ mW, $T_{bath} = 77$ K, $V_{sd} = 0.233$ V, $P_{MW} = 1$ W.

We now focus the attention on the revealed time scales. The component $I_{hot}$ shows two characteristic decay constants $\tau_1$ and $\tau_2$ (Fig. 4a, supplementary Fig. S2) ranging between a few picoseconds up to nanoseconds. The shorter time constant $\tau_1$ is approximately 40 ps and reflects the heat coupling of the graphene sheet to the substrate, since this time constant is also detected in pristine graphene without nanodiamonds (supplementary Fig. S3). In the regions away from the contacts, $\tau_1$ also comprises the NRET dynamics, because here, the signal to noise ratio limits the detection of $I_{sampling}$ to shorter time scales (see Fig. 2b, and supplementary Fig. S3). Most strikingly, $\tau_2$ significantly exceeds the decay times observed in pristine graphene (supplementary Fig. S3)[8]. To understand the longer decay times for the sample with nanocrystals, we need to discuss the dynamics of the NRET. It has been recently established, both theoretically and experimentally[2–4,6,19] that the near field interaction between optical emitters and graphene leads to a dramatic reduction of the lifetime of the emitters depending on the distance separating the emitter and the graphene sheet. Specifically, the lifetime $\tau_g$ of the emitter in the presence of graphene decreases as[3,4,19]

$$\frac{\tau_0}{\tau_g} = 1 + \frac{9\nu\alpha}{256\,\pi^3(\epsilon+1)^2}\left(\frac{\lambda_0}{d}\right)^4$$

(1)

where $\tau_0$ is the natural lifetime of the emitter, $\alpha$ is the fine structure constant, $\varepsilon$ is the permittivity of the sapphire, $d$ is the graphene-emitter distance, $\lambda_0$

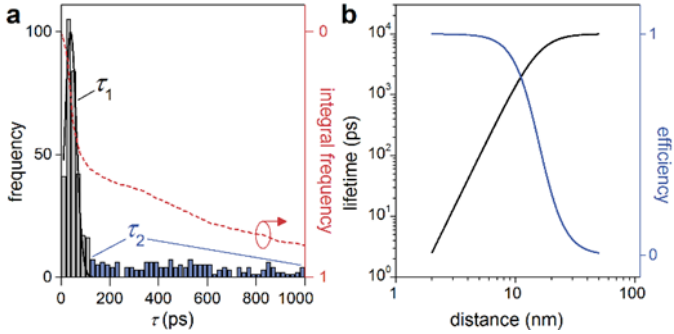

**Figure 4 Time-scales of nonradiative read-out of NV centers. a,** Frequency of prevailing time-constants for all positions on the graphene sheet with nanocrystals on top with contributions $\tau_1 \approx 40$ ps and $\tau_2$ up to nanoseconds. The dashed line is the integrated frequency over all delay times. **b,** Dependence of the lifetime on the distance between the NV and the graphene following Eq. 1. The lifetime is reduced by several orders of magnitude at short distances. Blue curve describes the efficiency of the NRET.

is the emission wavelength and $v = 1$ (2), if the emitter dipole is oriented parallel (perpendicular) to the graphene plane. For NV centers on a sapphire substrate ($\varepsilon = 3.12$), with an emission wavelength at 637 nm (zero phonon line)[21] and a natural lifetime of 10 ns, Eq. 1 results in a lifetime reduction from 10 ns at 50 nm to 65 ps at 5 nm (Fig. 4b). The physical mechanism of the NRET between an emitter and graphene is similar to the Förster resonant energy transfer (FRET) between two coupled emitters through a near-field dipole-dipole interaction, but contrary to FRET, the energy from the excited emitter is irreversibly transferred to the electronic excitations in graphene. The efficiency of the NRET is close to 100% for $d \leq 10$ nm (Fig. 4b). Comparing time scales, for $d > 50$ nm, the long 10 ns lifetime of NV centers is commensurate with the repetition frequency of the laser $1/f_{laser} = (76\text{ MHz})^{-1} \approx 13$ ns. This explains the role that $I_{offset}$ plays in the electronic readout of the NRET. For $I_{hot}$, the decay times correspond to the reduction of the lifetime of the NV centers due to NRET. The different characteristic times for this decay reflect the distribution of lifetimes of the numerous NV centers with respect to their distance $d$. Indeed, due to the size of the crystals (100 nm diameter) it is expected that the NRET is strongest for the NVs closest to the graphene sheet, leading to decay times in the ps regime (supplementary

Fig. S3), while the NVs further away contribute less to the current (Fig. 4b). Therefore, the long decay times observed in $I_{hot}$ (compared to pristine graphene without crystals) reflect the NRET-decay rate of NVs close to graphene. Since the measured characteristic times of $I_{hot}$ are significantly shorter than the natural lifetime of the NV centers, the detected long-lived currents stem from a NRET process and not from NV-fluorescence. Hereby, the measured current can be understood by extra electron-hole pairs in the graphene, generated at a time rate which corresponds to the lifetime of the NV centers modified by the NRET. Further verification of the electronic readout of the NV centers is given by the $P_{MW}$ dependence and the $P_{laser}$ dependence of the signal which are consistent with conventional optical detection of NV centers (supplementary Fig. S4).

**Conclusion:**

Our results show that the strong near-field interaction between graphene and optical emitters such as NVs in diamond can be harnessed to electronically readout the excited state of the emitters through NRET into electron-hole pair excitations in the graphene. The Auston switch technology enables the electronic ultrafast detection (in time scales shorter than the natural lifetime of the NV centers) of these electronic excitations which would otherwise be lost using slower, conventional electronics. To validate our conclusions, we induce electron spin resonance in the NV centers with resonant microwaves. Contrary to conventional fluorescence measurements utilised to detect the spin state of the NV centers, our novel approach enables us to detect the ESR signal electronically, giving access to characteristic lifetimes ranging from the ps regime (for emitters closest to the graphene sheet) up to the natural lifetimes of NV centers in the ns regime (for emitters further away from the graphene). These results pave the way for alternative architectures of quantum technologies and metrology that combine the advantages of optical emitters, such as robustness, long coherence times at room temperature, and sensitivity, with on-chip integration into electronic circuits.

**Methods**

**Fabrication of graphene.** Graphene films were grown by chemical vapor deposition (CVD) on copper foil in a hot-wall reactor. After pre-annealing the copper foil at 1000°C under a flow of 28 sccm $H_2$ for 40 min, graphene was grown for 30 min under a flow of 3.5 sccm $CH_4$ and 16 sccm $H_2$ at a total pressure of 10 mbar. Then, the copper-graphene foil was cooled to room temperature under growth atmosphere. For transfer, samples were spin-coated with poly(methylmethacrylate) (PMMA) and floated on 0.5 M aqueous $FeCl_3$ solution overnight. After complete dissolution of the copper foil, the films were rinsed with deionized water and transferred to the pre-structured sapphire substrates with a stripline circuit on top.

**Design of the stripline circuit.** Sapphire with a thickness of 430 μm, covered with a 300 nm thick silicon layer, is used as a substrate. The silicon is implanted with $O_2$ ions to yield an excess carrier lifetime of ≈ 1 ps. The silicon is etched in a first lithographic step to define the ultra-fast photo switch (Auston switch), using HF/ $HNO_3$ as etchant. Ti (5 nm) and Au (25 nm) are evaporated to form the striplines and the readout contact of the Auston switch. The two striplines are 15 μm distanced and they have a width of 5 μm. Each stripline has a total length of about 4.4 cm. The graphene is deposited on top of the striplines and defined to a 25 μm wide stripe in a last lithographic step by using $O_2$ plasma to remove the spare graphene. The distance between the graphene and the Auston switch is ≈ 250 μm. The nanodiamonds have a diameter of ≈ 100 nm and each comprises ≈ 500 NVs (Adamas Nanotechnologies, Inc., Raleigh, NC 27617). They are spin-coated onto the circuit from a propanol solution.

**On chip, time-domain THz spectroscopy.** A pulsed titanium:sapphire laser (repetition rate 76 MHz, pulse length 160 fs), operated at 811 nm, is used for the probe-pulse of the time-resolved measurements. The light is additionally converted by a non-linear fibre to a wavelength of λ = (535 ± 5) nm, and focused as the pump-laser onto the graphene with a 10x objective. Each pump-laser pulse excites both graphene and the NV centers. The ultrafast currents in the graphene-sheet are collected by the striplines. Consequently an electromagnetic transient, which is proportional to the initial current, will propagate along the striplines. After a certain time delay $\Delta t$, the probe-pulse triggers the Auston-switch and the presence of the electromagnetic transient drives an electric current, which decays within 1 ps. This current is read out as $I_{sampling}(\Delta t)$ and it is proportional to the electric field of the propagating electromagnetic transient. The moment, when the pump-pulse hits the graphene, defines $\Delta t \approx 0$ ps. Both laser beams are focused through the same objective on the sample, but the position of the pump beam on the graphene-sheet can be independently scanned by a motorized mirror. The spot size of the pump beam is 3- 4 μm. This beam has a power of $P_{laser}$ = 0.3- 3 mW. The probe laser power is set to 100 mW. All laser powers are measured in front of the objective. We use a dual frequency modulation of the pump- and the probe-beam, and read out the pre-amplified electric current $I_{sampling}$ with a lock-in amplifier. The striplines are either connected to a voltage source ($V_{sd}$) or connected via a bias tee to both an amplified microwave source with power $P_{MW}$ and the voltage source for electron spin resonance (ESR) experiments. All measurements are done in vacuum ($10^{-5}$ mbar) at $T_{bath}$ = 77 K.

**Acknowledgements**

We gratefully thank Antoine Reserbat-Plantey for technical support. This work was supported by the ERC Grant NanoREAL (n°306754) and the "Center of NanoScience (CeNS)" in Munich. L.G. acknowledges financial support from Marie-Curie International Fellowship COFUND and ICFOnest program. F.K. acknowledges support by the Fundacio Cellex Barcelona, the ERC Career integration grant 294056 (GRANOP) and the ERC starting grant 307806 (CarbonLight).

**Author Contribution**

A.B. and L.G. performed the experiments and analyzed the data together with A.W.H, F.K., M.S., J.A.G., M.S.B., H.H., and H.K.. F.K. and A.W.H. conceived the study. All authors co-wrote the paper.

# Supplementary Information

Ultrafast electronic detection of diamond NV centers coupled to graphene

Andreas Brenneis, Louis Gaudreau, Max Seifert, Helmut Karl, Martin S. Brandt, Hans Huebl, Jose A. Garrido, Frank H.L. Koppens*, Alexander W. Holleitner*

**Supplementary Notes**

1. Ultrafast time-resolved photocurrents in graphene with nanocrystals on top:
The NRET generates additional electron-hole pairs in the graphene at a time-constant which corresponds to the lifetime of the NV centers modified by the NRET. The dynamics of the additional electron-hole pairs in graphene are governed by the same physical mechanisms as of hot, photo-excited electron-hole pairs. Such photocurrents are actively discussed either in terms of the so-called photothermoelectric effect [8,13–14,16] or built-in electric fields created by the different work function mismatches between the different materials (diamond-graphene, metal-graphene)[8,15,16,36]. In the following, we characterize the NRET-currents in detail.

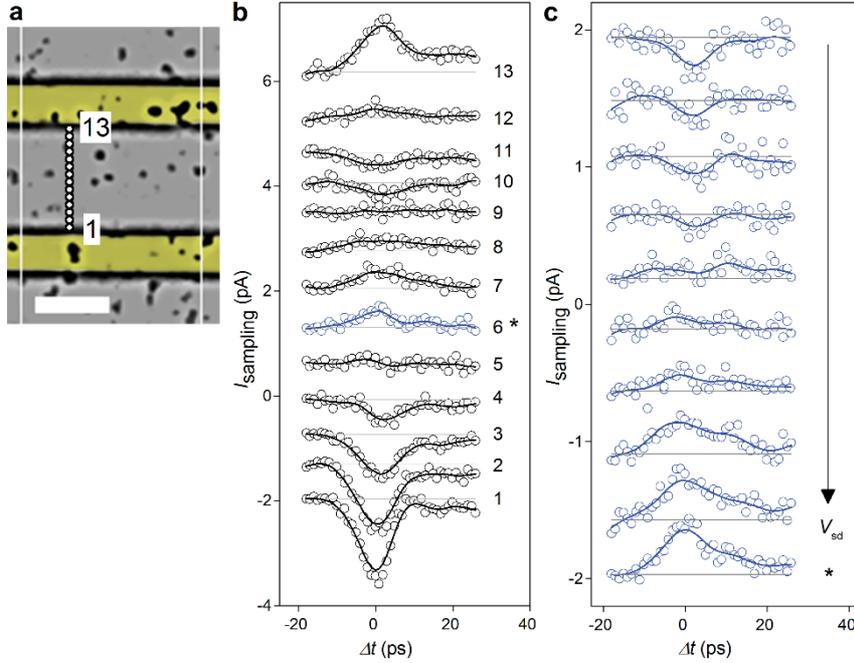

**Fig. S1 Time-resolved currents in graphene with nanocrystals on top. a**, Microscope image of a graphene-sheet with nanocrystals on top defining the excitation positions from 1, 2, 3, …, to 13. Scale bar, 10 µm. **b**, Time-resolved current $I_{sampling}$ for positions 1- 13. An offset is added to the data for clarity. **c**, The trace at position 6 as a function of an external bias voltage. The bias is increased in steps of 17 mV. The data in Fig. S1b and Fig. 2 of the main manuscript were measured under the same bias as the trace highlighted by *. In Fig. S1c, no offset is added to the data, i.e. the background current $I_{offset}$ depends linearly on the applied bias. All lines are guidelines to the eye. The experimental parameters are $\lambda = 535$ nm, $P_{laser} = 2.0$ mW, $T_{bath} = 77$ K, $f_{MW} = 2.800$ GHz, $P_{MW} = 0.8$ W.

Fig. S1a highlights positions 1 to 13 of the laser focused on the discussed graphene sheet with nanocrystals on top. For each position, we measure $I_{sampling}$ vs. $\Delta t$ (supplementary Fig. S1b). Close to the metal striplines (positions 1 and 13) a dominant displacement current is observed as a first peak of $I_{sampling}$ at $\Delta t \approx 0$ ps. This ultrafast displacement current is caused by photogenerated charge carriers which screen the local electric fields[7,8]. The ultrafast displacement current has a negative (positive) amplitude at position 1 (position 13) consistent with the direction of the local built-in electric fields at the metal-graphene (graphene-metal)-interface[8]. The displacement current is sensitive to built-in potentials caused by the work function difference of graphene and the metal. From position 1 to 13 $I_{sampling}$ comprises a rich structure and the ultrafast displacement current at $\Delta t \approx 0$ ps changes the sign twice. The behavior of the displacement current can be explained by the presence of local electric fields caused by the different work functions of graphene and the diamond nanocrystals. We prove this by the application of an external bias voltage and, therefore, a macroscopic electric field which flattens the local field distribution. To this end, the trace at position 6 is highlighted in blue in Fig. S1b. When an external bias is applied (supplementary Fig. S1c), the displacement peak at $\Delta t \approx 0$ ps changes sign according to the bias voltage. In other words, for positions far away from the contacts, one needs an external electric field to observe and enhance a time-resolved photocurrent. We note that at the positions of the nanocrystals and for a finite bias voltage, we also observe the long-lived photocurrents of $I_{sampling}$ (Fig. 2 of the main manuscript), as discussed in the next section.

2. Comparison of the time-resolved, long-lived photocurrents with and without NVs:
The supplementary Fig. S2 depicts the data of Fig. 2c of the main manuscript again. As in the main manuscript, the blue line is a fit to the data with two exponential Gaussians with decay times $\tau_1$ and $\tau_2$. For all samples with and without nanocrystals, we observe two decay times of $I_{sampling}$. In the following, we establish the physical meaning of these. For clarity, we distinguish $\tau_1^{NVC}$ and $\tau_2^{NVC}$ for samples with nanocrystals (and NVs) from $\tau_1^{pristine}$ and $\tau_2^{pristine}$ for pristine samples without nanocrystals.
We perform measurements of $I_{sampling}$ for all positions on the graphene sheets with nanocrystals on top (supplementary Fig. S3a) and without (supplementary Fig. S3e) with equivalent experimental parameters. Figures S3b and 3c show the lateral distribution of $\tau_1^{NVC}$ and $\tau_2^{NVC}$, and the Figures S3f and 3g show the lateral distribution of $\tau_1^{pristine}$ and $\tau_2^{pristine}$, respectively. Most importantly, $\tau_1^{NVC}$ shows up at positions in the middle of the sample where nanocrystals are deposited (supplementary Fig. S3b, circles and triangles in Fig. S3a as well as Fig. 2a of the main manuscript).

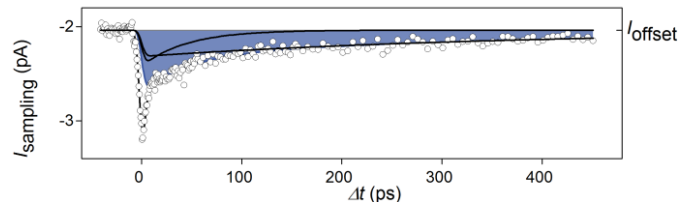

**Fig. S2. Ultrafast electronic read-out of NV centers and fitting curves**. Time-resolved current $I_{sampling}$ reprinted from Fig. 2c of the main manuscript with two exponential Gaussians as fitting curves (black lines). The blue line and area represent the sum of the two fitting curves.

2.1 The timescales $\tau_1^{NVC}$ and $\tau_1^{pristine}$:
The time-constant of $\tau_1^{NVC}$ [$\tau_1^{pristine}$] has a maximum (39 ± 27) ps [(43 ± 14) ps] (Figs. S3d and S3h). Both timescales are identical, which substantiates the argument that the transferred charge carriers (from the NV center to the graphene) are governed by the same physical mechanisms as for hot photo-excited charge carriers. The hot electrons contribute to the current as long as the electron reservoir is in non-equilibrium. In a simple model, we estimate the temperature $T$ of the photo-excited graphene caused by the interfacial heat flow to the underlying $Al_2O_3$-substrate. The change of $T$ is described by the differential equation[39]

$$\frac{dT}{dt} = \frac{P_{cooling}}{C_{graphene}A} = -\frac{(T-T_0)G_{Kapitza}A}{C_{graphene}A} = -(T-T_0)\kappa \quad , \qquad S(1)$$

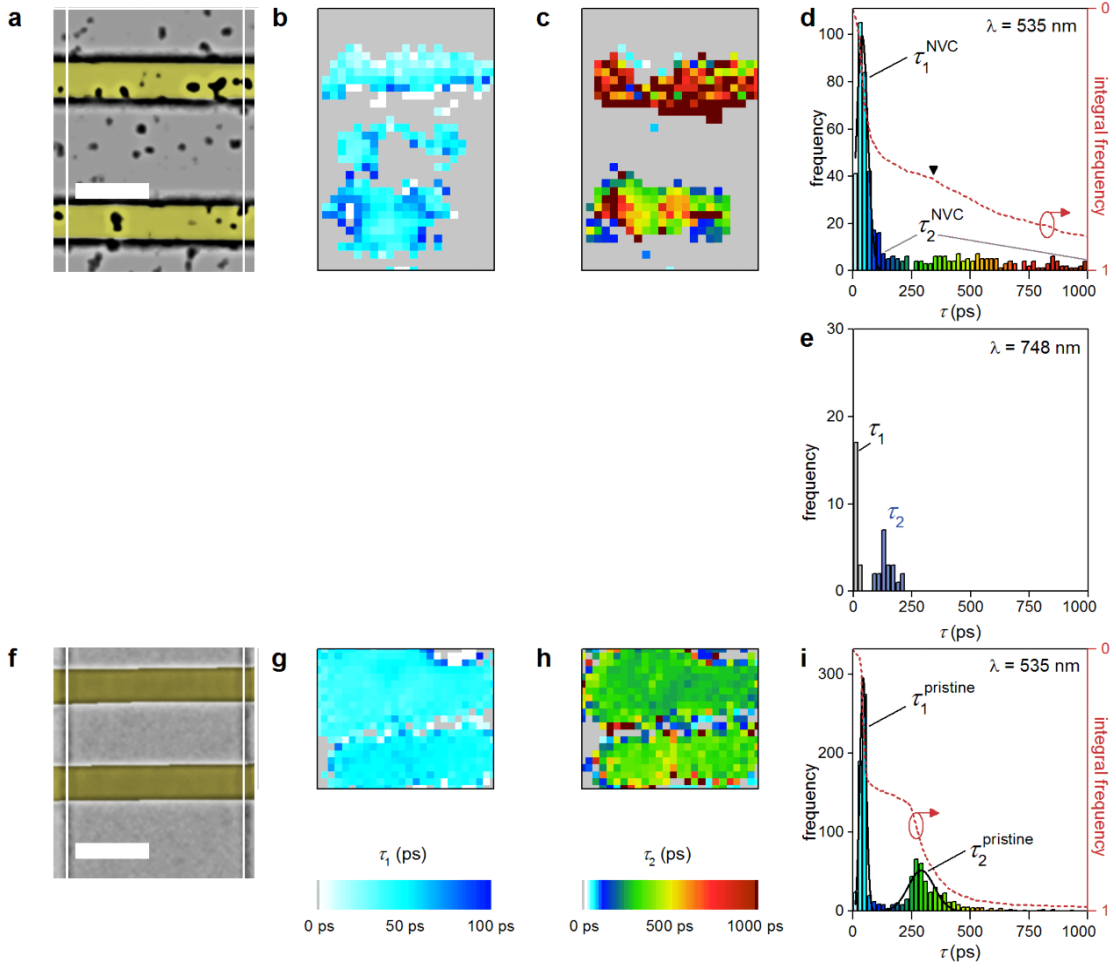

**Fig. S3 Time-resolved currents in graphene with and without NVs. a,** Microscope image of the discussed graphene-sheet with nanocrystals on top. Scale bar, 10 µm. **b,** The spatial distribution of the fitted $\tau_1^{NVC}$ for the data as presented in Fig. 2 of the main manuscript. Fitting curves are presented in Fig. S2. The spatial dimensions are the same as in a. The experimental parameters are $\lambda = 535$ nm, $P_{laser} = 3.0$ mW, $T_{bath} = 77$ K. **c,** Corresponding spatial distribution of $\tau_2^{NVC}$. **d,** Frequency and integrated frequency of prevailing time-constants for all positions on the graphene sheet with nanocrystals on top with a maximum at $\tau_1^{NVC} = (39 \pm 27)$ ps. **e,** Frequency of prevailing time-constants for all positions on the graphene sheet with nanocrystals on top with a maximum at $\tau_2^{NVC} \approx 200$ ps. For this measurement, the laser wavelength is 748 nm, i.e the optical excitation is below the optical transition energy of NV centers. **f,** Microscope image of a pristine graphene-sheet without nanocrystals. Scale bar, 10 µm. **g,** Spatial distribution of $\tau_1^{pristine}$ with same lateral dimension as in e. The experimental parameters are $\lambda = 535$ nm, $P_{laser} = 3.0$ mW, $T_{bath} = 77$ K. **h,** Corresponding spatial distribution of $\tau_2^{pristine}$. **i,** Frequency and integrated frequency of prevailing time-constants for all positions on the graphene sheet without nanocrystals on top with a maximum at $\tau_1^{pristine} = [(43 \pm 14)$ ps$]$ and $\tau_2^{pristine} \approx 260$ ps. See supplementary text for discussion.

with $P_{cooling}$ the cooling power, $C_{graphene}$ the heat capacitance of graphene, $T_0$ the temperature of the substrate, $G_{Kapitza}$ the Kapitza conductance between the graphene and the substrate, $A$ the interfacial area, and $\kappa = G_{Kapitza} / C_{graphene}$. Equation S(1) can be easily solved by an exponential function with a time-constant $\kappa^{-1}$. For $C_{graphene} = 1.5 \times 10^{-15}$ JK$^{-1}$µm$^{-2}$ and $G_{Kapitza} = 4 \times 10^{-5}$ WK$^{-1}$µm$^{-2}$, $\kappa^{-1}$ equals 38 ps[39]. Hereby, we explain the obtained timescale of $\approx 38$ ps seen on pristine graphene and on graphene with nanocrystals by a Kapitza conductance at the graphene/Al$_2$O$_3$-interface of $4 \times 10^{-5}$ WK$^{-1}$µm$^{-2}$, which is a very reasonable value[39].

During this cooling time and at the presence of an electric field (see supplementary Fig. S1), the hot electrons contribute to the time-dependent $I_{sampling}$. This is valid for a finite bias at positions in the middle of the sample where nanocrystals are deposited (circles and triangles in supplementary Fig. S3a and Fig. 2a of the main manuscript), and independent of the bias, at the metal contacts where built-in electric fields give rise to the electric field (supplementary Figs. S3b and S3f).

2.2 The timescales $\tau_2^{\text{pristine}}$ and $\tau_2^{\text{NVC}}$:
Y. K. Koh et al. demonstrated that a lateral heat flow within graphene is only dominant for a lateral distance of ≈ 180 nm to the metal striplines[40]. On the other hand, the phonon temperature of the metal striplines decays on a timescale of hundreds of picoseconds through transverse heat diffusion and coupling to the substrate[41,42]. In turn, there are hot electrons in the vicinity of the metal-graphene interface, and $I_{\text{sampling}}$ shows a decaying component close to the metal contact.
Supplementary Fig. S3h demonstrates that $\tau_2^{\text{pristine}} \approx 260$ ps, and Fig. S3g verifies that $\tau_2^{\text{pristine}}$ is rather homogeneous along the metal-graphene interfaces. Both results corroborate the interpretation of an additional heat-sink caused by the metal contacts. We point out that this interpretation is consistent with independent measurements on freely suspended, pristine graphene, in which we observed an exponential timescale in the range of 100 to 300 ps at the graphene-metal interface[8].

Intriguingly, $\tau_2^{\text{NVC}}$ shows longer timescales up to nanoseconds (supplementary Fig. S3d), and the lateral distribution of $\tau_2^{\text{NVC}}$ is less homogeneous (supplementary Fig. S3c). We still detect a tiny bump in the integrated frequency for $\tau_2^{\text{NVC}}$ (triangle in supplementary Fig. S3d), which we tentatively interpret to be a reminiscence of the lateral heat flow to the metal contacts. However, there is a dominating NRET-process with distinctively longer $\tau_2^{\text{NVC}}$ in close vicinity of the metal striplines (supplementary Fig. S3c). The vicinity to the striplines helps to measure $\tau_2^{\text{NVC}}$. There, the strong static interfacial built-in electric fields make the hot electrons propagate; independent of the time-delay to the pump-pulse.
To conclude this section, time-constants exceeding ≈ 260 ps are only seen for graphene with nanocrystals on top, and they are significantly shorter than the natural, characteristic lifetime of NV centers without the NRET. Since the corresponding currents can be controlled by an electron spin resonance, the underlying processes are related to the optically excited spins in the NV centers; proving the interpretation of a non-radiative energy transfer from the NV centers to the graphene.

3. Electron Spin Resonance (ESR) Measurements:
To verify the influence of the NV centers on the time-resolved current, a microwave source is applied to the stripline to induce an electron spin resonance (ESR) of the ground state of the NV center. The ESR analysis of the current (Figures 3c and 3d of the main manuscript) is performed as follows: the offset amplitude of $I_{\text{sampling}}$ is measured as the microwave frequency $f_{\text{MW}}$ is swept. Then, the frequency-dependent data are offset-corrected by a linear fit, and alias effects at multiples of the laser repetition frequency are subtracted using Gaussian peak functions. The resonance is accounted by a Lorentzian fit and the corrected data is labeled $I^*_{\text{offset}}$ in the plots.
The resonance dip is consistent with a saturated ESR which reduces the occupation of the optically excited state with $m_s = 0$, which in turn, reduces the NRET. Further proof that the observed microwave frequency dependence of $I_{\text{offset}}$ originates from the ESR within the NVs, can be obtained by studying the broadening of the peak with pump laser intensity and microwave signal power. The laser power broadening of the ESR peak is depicted in Fig. S4a for different laser intensities, fitted to a Lorentzian function to obtain the full width half maximum (FWHM). The data show a clear linear increase in the FWHM with laser intensity (supplementary Fig. S4b). Fig. S4c illustrates how the resonance line is broadened by increasing microwave power $P_{\text{MW}}$. Within the experimental error, the FWHM scales with the amplitude of the driving signal, i.e. with $P_{\text{MW}}^{1/2}$ (supplementary Fig. S4d). Essentially, the NRET-current depends on both the microwave and the pump laser.

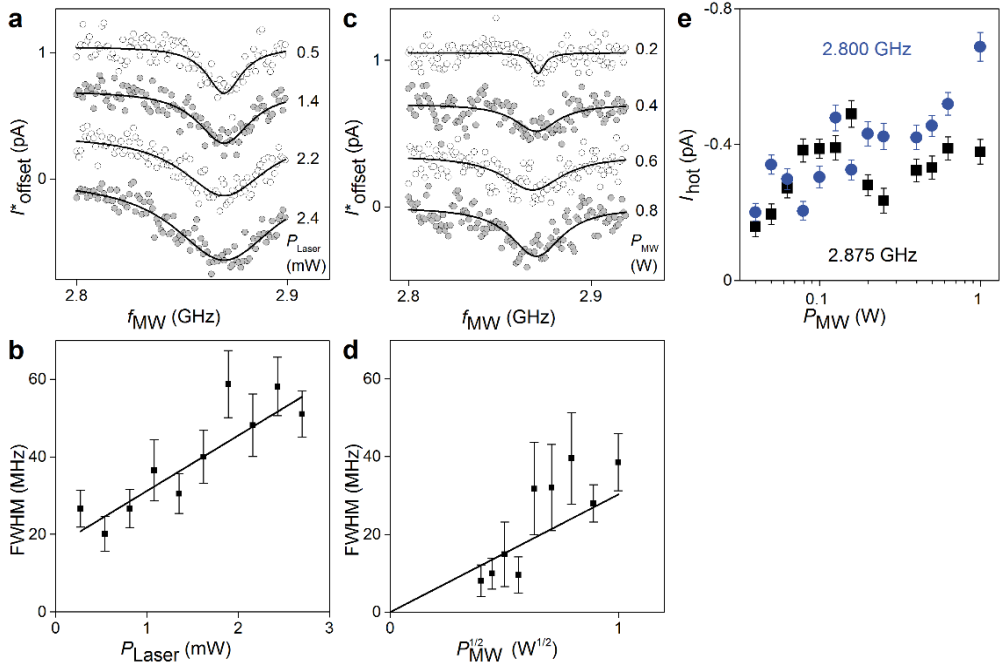

**Fig. S4. Power dependence of ESR. a,** Laser power dependence of $I^*_{offset}$ vs. $f_{MW}$ with Lorentzian fitting lines. **b,** Extracted FWHM vs. $P_{Laser}$. Line is a linear fit. **c,** and **d,** microwave power dependence of $I^*_{offset}$ vs. $f_{MW}$ with fitting lines. The experimental parameters are $\lambda = 535$ nm, $T_{bath} = 77$ K, $V_{sd} = 0$ mV, (a,b) $P_{MW} = 1$ W, (c,d) $P_{laser} = 2.0$ mW **e,** The amplitude of the ultrafast photocurrent $I_{hot}$ is reduced under resonant microwave excitation (2.875 GHz) compared to the non-resonant excitation (2.800 GHz) for microwave powers above 0.2 W measured at the position of the dotted circle in Fig. 2a. The experimental parameters are $\lambda = 535$ nm, $T_{bath} = 77$ K, $V_{sd} = 0$ mV, $P_{laser} = 2.0$ mW.

4. Power dependence of the ESR measured by $I_{hot}$:

Because the amplitude of the component $I_{hot}$ is 0.3-0.4 pA at an overall noise floor of 0.1 pA, the detection of the ESR via $I_{hot}$ with an ESR-dip of about 0.2 pA is inherently close to the experimental resolution (compare Figs. 3a and 3b of the main manuscript). To still demonstrate the power dependence of the ESR-dip also in $I_{hot}$, we measure $I_{hot}$ for a non-resonant and resonant ESR-microwave-excitation. In the supplementary Fig. S4e, the squares (circles) depict $I_{hot}$ for $f_{MW} = 2.875$ GHz ($f_{MW} = 2.800$ GHz) vs. the power of the applied microwave, which corresponds to a resonant (non-resonant) electron spin-excitation. Within the experimental error, the power-dependent increase of the ESR-dip is clearly resolvable.

5. Correlation of the microscope image to PL intensity of the NV centers:

Fig. S5 shows the correlation of the microscope image (main manuscript Fig. 2a) to the photoluminescence intensity measured on the presented sample. For all positions of nanodiamond clusters on the sample (circles and triangle in Fig. 2a of the main manuscript) the observed PL emission correlates to the black dots of the microscope image.

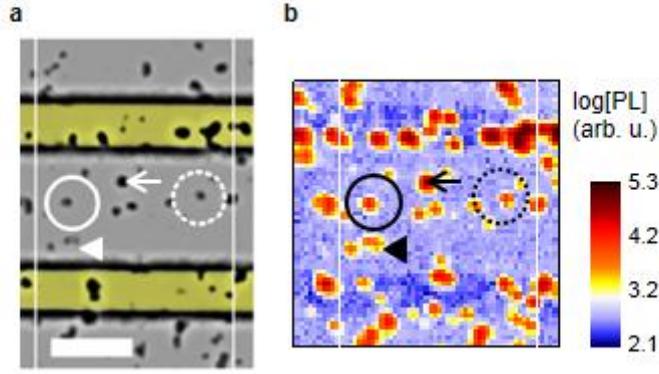

**Fig. S5. Correlation of the microscope image to the photoluminescence intensity of the NV centers. a,** The nanodiamond clusters observed in the microscope image as black dots can be correlated to the photoluminescence signal of the sample, presented in **b,**. The dotted circle denotes the excitation position of Fig. 3a, and Fig. 3b. The triangle denotes the excitation position of Fig. 3d. The arrow indicates the excitation position of Fig. 3c, Fig. 3e, and 3f of the main manuscript.

6. Photoluminescence spectrum of the NV centers:
Fig. S6 shows a PL spectrum taken at the position marked by the black arrow in Supplementary Fig. S5b. The peaks at the zero phonon line (ZPL) of $NV^-$ and $NV^0$ are highlighted.

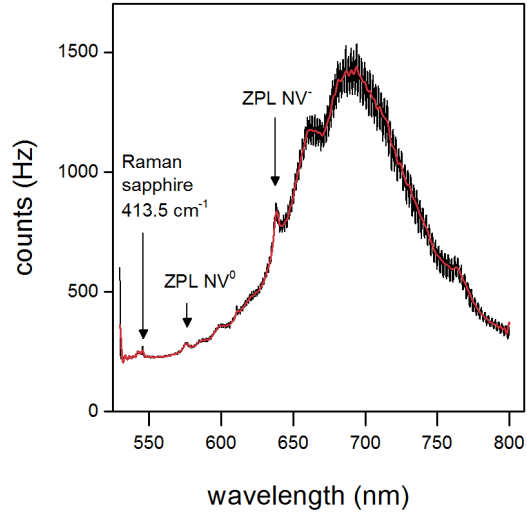

**Fig. S6. Photoluminescence spectrum of the NV centers.** The spectrum is taken at the position marked by the black arrow in Fig. S5 (excitation wavelength 532 nm, $P_{laser}$ = 0.33 mW). Raw data is shown as black line. A low pass filter (red line) is applied to remove the interference fringes that stem from reflections in the CCD detector. The zero phonon lines (ZPL) of the NV center and the Raman shift from the sapphire substrate are marked in the spectrum.

7. NV PL lifetimes:
Fig. S7 shows a time resolved photoluminescence histogram for NV centers located on the sapphire substrate and on the graphene, respectively. The data are fitted by two exponential Gaussians. On sapphire and on graphene, we observe a component on the time scale of $\tau_b \approx 10$ ns, which is assigned to the lifetime of the radiative recombination. Only for NV centers located on the graphene, a fast component $\tau_a$ in the order of hundreds of picoseconds is observed. This quenching is caused by an energy-transfer to the graphene.

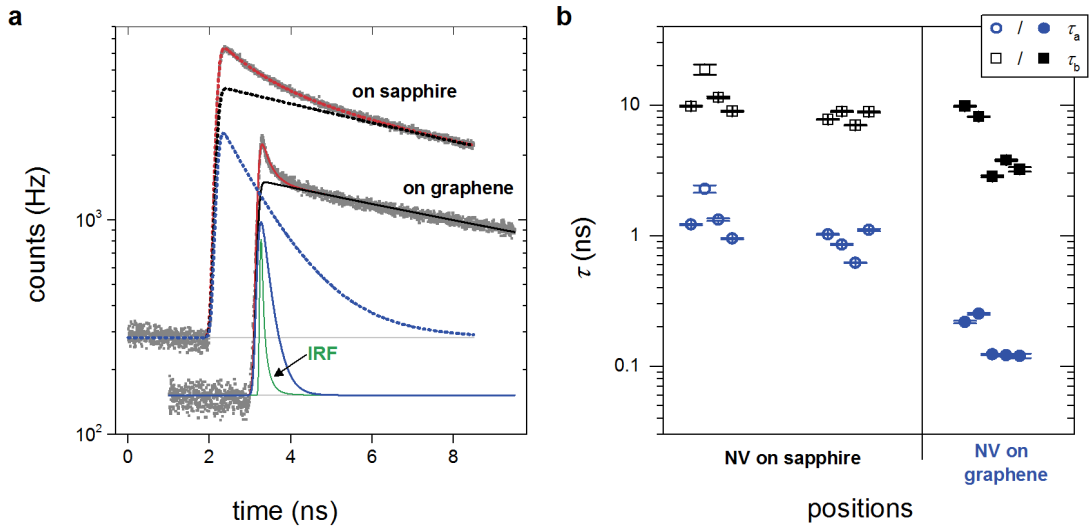

**Fig. S7. Photoluminescence lifetimes of the NV centers. a,** Time-resolved photoluminescence histogram for NV centers located on the sapphire substrate and on the graphene. The data are fitted by two exponential Gaussians ($\tau_a$ and $\tau_b$, blue and black line, respectively). The instrument response function (IRF) is shown as comparison. **b,** The component $\tau_1$ is faster for all positions on the graphene than on the sapphire substrate.